\documentclass[aps,prd,groupedaddress]{revtex4-1}

\usepackage{dsfont}
\usepackage{natbib}
\usepackage{amssymb}
\usepackage{graphicx}
\usepackage{amsmath}
\usepackage{url}
\usepackage{epstopdf}
\usepackage{color}
\usepackage{enumerate}

\newcommand{\ket}[1]{|#1\rangle}
\newcommand{\braket}[2]{\langle #1|#2\rangle}

\renewcommand{\H}{\mathcal{H}}
\newcommand{\Hp}{\mathcal{H}_{\text{p}}}
\newcommand{\Mp}{M_\text{Pl}}
\newcommand{\M}{M_\star}
\newcommand{\di}{\partial}
\newcommand{\pp}{p_{\phi}}
\newcommand{\pv}{p_{\varphi}}
\newcommand{\vi}{\varphi}
\newcommand{\Ld}{\mathcal{L}_D}
\newcommand{\Hd}{\mathcal{H}_D}

\newcommand{\sq}{\sqrt{q}}
\newcommand{\p}{\partial}

\def\be {\begin{equation}}                                                                            
\def\ee  {\end{equation}}                                                                             
\def\bea {\begin{eqnarray}}                                                                            
\def\eea {\end{eqnarray}}                                                                            
\def\nn {\nonumber}

\begin{document}

\title{Semiclassical cosmology with polymer matter}

\author{Syed Moeez Hassan}
\email{shassan@unb.ca}
\affiliation{Department of Mathematics and Statistics, University of New Brunswick, Fredericton, NB, Canada E3B 5A3} 

\author{Viqar Husain}
\email{vhusain@unb.ca}
\affiliation{Department of Mathematics and Statistics, University of New Brunswick, Fredericton, NB, Canada E3B 5A3}

\begin{abstract}

In loop quantum cosmology, polymer quantization is applied to gravity and Schrodinger quantization to matter. This  approach misses interesting cosmological dynamics coming from the polymer quantization of matter. We demonstrate this in semiclassical cosmology with a scalar field and pressureless dust: gravity is kept classical, dust is used to fix the time gauge, and polymer quantization effects are isolated in the scalar field. The resulting dynamics shows a period of inflation, both with and without a scalar potential, and the emergence of a classical universe at late times. Since gravity is not quantized, the cosmological singularity is not resolved, but our results suggest that polymer quantization of both gravity and matter are important for  a complete picture.

\end{abstract}

\date{\today}
\maketitle

\section{Introduction}

Inflation is a part of the standard model of cosmology \cite{Baumann}. It explains key observations in our universe including large scale spatial homogeneity and structure formation. It is natural to ask what role, if any, quantum gravity effects play in explaining or  modifying the standard inflationary picture, and whether signals of quantum gravity are writ large on the sky \cite{Easther:2001,Shankaranarayanan:2002ax}. This question continues to be discussed from various points of view including  string theory \cite{stringcosm}, non-commutative geometry \cite{Brandenberger:2002nq,NCGcosm}, and loop quantum gravity (LQG) \cite{lqc-rev3, Ashtekar:2011ni,lqc-rev1} Of these varied approaches, most of the work related to cosmology is from the first and last.  

One of the main lessons  emerging from LQG is a method of quantization called polymer quantization. The key feature of this is the use of a non-separable Hilbert space that has a built in notion of fundamental discreteness.    Although this method can be used for quantization of  any classical system, the most detailed studies have been to cosmological models written in the connection-triad  variables for canonical gravity, an area called loop quantum cosmology (LQC). Perhaps the most important result  is a general mechanism for singularity resolution in a variety of models \cite{Ashtekar:2006uz,lqc-robust}. This comes about from the operator realization of the  Hamiltonian constraint on this Hilbert space with the scalar field (with zero potential) used as a clock.   

However all models studied to date in LQC use the usual Schrodinger quantization prescription applied to matter, and polymer quantization applied to  the gravitational degrees of freedom. This is not natural from the point of view of full LQG, where polymer quantization is taken to be a fundamental ingredient, and is applied to both gravity and matter sectors \cite{Thiemann:1997rt}. Indeed, given this starting point, it is natural to expect that in a semiclassical regime, there will first be an emergence of a ``polymer quantum field theory" (PQFT) on curved spacetime, and a subsequent lower energy emergence of usual QFT on curved spacetime. These are open issues, and so far there has been some preliminary investigations of the polymer quantum field theory on flat spacetime with some interesting physical results \cite{vh-ak,poly-prop,poly-detector}. 

In this paper we investigate this intermediate regime as it applies to cosmology.  We treat gravity classically and study the semiclassical Hamiltonian theory with polymer quantized matter in a Gaussian state.   We do this in the Arnowitt-Deser-Misner (ADM) variables, with the understanding that if gravity is kept classical, there is little difference from connection-triad  variables. Our aim is to isolate the effects of polymer quantization of matter on cosmological dynamics as a prelude to a more complete treatment. 

This approach has been studied  before for homogeneous and isotropic universe with a massless and massive scalar field \cite{polycosm1,polycosm2}. We extend these works by adding a dust field which introduces an extra degree of freedom in the system. The dust is used to fix a time gauge \cite{dust-time,dust-time-cosm}, and the corresponding physical Hamiltonian is used to study the dynamics. Without a ``solution" to the problem of time, this time gauge provides a potentially useful and relatively simple physical Hamiltonian for studying semiclassical and quantum gravity.  

Our main result is that polymer quantization of matter in this model naturally produces a period of inflation followed by a graceful exit into a classical regime. This occurs both with and without a scalar field potential. These results are qualitatively similar in some aspects, but quantitatively quite different from the above works. 

In the next section (II) we review the Hamiltonian theory in dust time gauge in a general setting, followed by a description of the polymer quantization for the scalar field in Sec. III. This quantization is different from that employed in LQG, but it captures the main feature that the scalar kinetic operator is written in a manner similar to that for the gravitational kinetic term in LQC. In Section IV we present the results of semiclassical dynamics, followed in Sec. V by a summary of results, and a discussion of possible further  developments. 

\section{Hamiltonian theory}\label{sec:setup}

The theory we study is general relativity with  a minimally coupled scalar field $\phi$ and a pressureless dust field $T$.  The dust Lagrangian on a manifold with metric $g_{ab}$  is 
 \be
\Ld(g,T,M) = -\dfrac{1}{2} M \sqrt{-g} \left( g^{ab} \p_aT \p_bT + 1 \right),
\ee
where $M$ is the field that enforces the constraint that the dust is timelike. The stress-energy tensor  is 
\be
T_{ab}=M \left( \p_aT \p_bT - \frac{1}{2} g_{ab} (1+\p_cT \p^cT) \right).
\ee
therefore on shell, $M$ is the dust energy density. 

The Hamiltonian theory is obtained by substituting the Arnowitt-Deser-Misner (ADM) parametrization of metric 
\be
ds^2 = -N^2dt^2 + (dx^a + N^a dt)(dx^b + N^b dt)q_{ab} \label{admm}
\ee  
into the Lagrangian. $ N$ and $ {N}^a$ are the lapse and shift fields and $q_{ab}$ is the spatial metric. The dust momentum is 
\be
p_T = \frac{\p \Ld}{\p \dot{T}} = M \frac{\sq}{N} (\dot{T} - N^a \p_a T), \label{pT}
\ee
where $q = \text{det}(q_{ab})$.  The dust   Hamiltonian density is  
\be
\Hd =  \frac{p_T^2}{2M \sq} + \frac{1}{2}M\sq\ (1+q^{ab}\p_aT \p_bT)
\ee
Variation of the canonical from of the action with respect to $M$ gives 
\be
M^2 = \frac{p_T^2}{q} \frac{1}{(q^{ab}\p_aT\p_bT + 1)}.
\ee
Substituting this back gives 
\be
\Hd = \text{sgn}(M)\  p_T \sqrt{1+q^{ab}\p_aT \p_bT}.  \label{HD}
\ee
The ADM canonical action of general relativity with massive scalar field and dust is   
\be
 S= \int d^3x\  dt  \ \left(\pi^{ab} \dot{q}_{ab}  + p_\phi \dot{\phi} + p_T \dot{T} - N{\cal H} - N^a {\cal C}_a   \right) , 
\ee
where
\begin{subequations} 
\bea
{\cal H} &\equiv&  {\cal H}_G + {\cal H}_\phi + \Hd,\nn\\
              &=& 
\frac{2}{\Mp^{2}}  \frac{ \pi^{ab}\pi_{ab} - \frac{1}{2} \pi^2 }{\sqrt{q}}  + \frac{\Mp^{2}}{2} \sqrt{q} (\Lambda -  \, {}^{(3)} \!R) \nn\\
   && +   \left(  \frac{p_\phi^2}{2\sqrt{q}} +  \frac{1}{2}  \sqrt{q}q^{ab} \di_a\phi \di_b \phi   + \frac{1}{2} m^2 \phi^2 \right)  + \Hd, \\	 
{\cal C}_a &\equiv&  -D_b \pi^b_{\ a} +  p_\phi \di_a\phi - p_T \di_a T,
\eea
\end{subequations}
$\Mp^{2} = 1/8\pi G$  and $D_{a}$ is the  spatial metric-compatible covariant derivative.  

\subsection{Dust time gauge}

We study this theory in the dust time, defined by  $T=\epsilon t$ with $\epsilon= \pm 1$. In the end we will make one choice, but we keep this freedom for now  to interpret physically the corresponding physical Hamiltonian.  The requirement  that the gauge be preserved in time requires 
\be
 \dot{T} = \epsilon = \{ T, \int d^3x\   N {\cal H} \}|_{T=t} = \text{sgn}(M) N. \label{gauge-pres}
\ee 
The physical Hamiltonian $\Hp$ is obtained by substituting the gauge into the dust symplectic term in the canonical action, which identifies  $\Hp = -\epsilon p_T$. Using (\ref{HD}) and solving the Hamiltonian constraint 
\be
{\cal H}_G +  {\cal H}_\phi + \text{sgn}(M) p_T = 0 
\ee
gives 
\be
\Hp = -\epsilon p_T  =  \text{sgn}(M) \ \epsilon \left( {\cal H}_G + {\cal H}_\phi \right) = N \left( {\cal H}_G + {\cal H}_\phi \right)\ ,
\ee
using  (\ref{gauge-pres}) for the last equality.  It is also useful to note, using (\ref{pT}) and (\ref{gauge-pres}), the relation
\be
  p_T = \epsilon  \sqrt{q}\  \frac{M}{N} = \epsilon \ \sqrt{q}\  \frac{\text{sgn}(M)}{N}\ |M| = \sqrt{q}\  |M| \ \ .    
\ee
which shows that  $p_T > 0$  for $M\ne 0$, and 
\be
\Hp = -\epsilon \sqrt{q}\  |M| = N \left( {\cal H}_G + {\cal H}_\phi \right).  
\ee  
We note also that the requirement that the dust Hamiltonian satisfy ${\cal H}_D = \text{sgn}(M)p_T\ge 0$ implies $\text{sgn}(M)=+1$, since $p_T= \sqrt{q}\ |M| \ge 0$. This means that the dust field satisfies the weak energy condition.
  With this choice (\ref{gauge-pres}) gives $N=\epsilon$.  In the following we make the choice $N=\epsilon = -1$ which gives the manifestly positive physical Hamiltonian density
\be
\Hp = \sqrt{q}\ |M| = -  \left( {\cal H}_G + {\cal H}_\phi \right) \ge 0. \label{Hp2}
\ee

\subsection{Semiclassical approximation for cosmology} 

The reduction to flat FRW cosmology is obtained  by setting 
\be
q_{ab} = a^2(t) e_{ab}, \quad \pi^{ab} = \frac{P_a}{6a(t) } e^{ab}, 
\ee
where $e_{ab}$ is the flat Euclidean metric. The (dust time) gauge fixed action is 
\begin{subequations}
\bea
S_\text{R} &=&V_0 \int dt \left(P_a\dot{a} +p_\phi \dot{\phi}  -  \Hp \right),\\
\Hp &\equiv& -\left({\cal H}_G + {\cal H}_\phi \right) \nn\\
&=&\frac{P_a^2}{12 a\Mp^{2}} - a^3  \Mp^2 \Lambda - \frac{p_\phi^2}{2a^3} - \frac{1}{2} a^3 m^2 \phi^2,    \label{Hdust}
\eea
\end{subequations}
where the last equation follows from (\ref{Hp2}), and $V_0=\int d^3x$ is a fiducial volume. Eqn. (\ref{Hp2}) also requires that initial data for this system be chosen such that $\Hp=e\ge 0$ for some constant $e$: this may be rearranged  as the Friedman equation   
\be
3 a^3 \left( \frac{\dot{a}}{a} \right)^2 = a^3 \Lambda + \frac{1}{\Mp^2} \left({\cal H}_\phi + e\right), 
\ee
indicating positive $e$ corresponds to positive dust energy density. 

%-------------------------------------------------%

Our goal in the following sections is to study this model in the semiclassical approximation. This is defined using a state  of the scalar field 
  $|\Psi\rangle(\bar{\phi},\bar{p}_\phi;\sigma)$ of width $\sigma$, peaked at  the phase space point $(\bar{\phi},\bar{p}_\phi)$. Such a state determines an effective Hamiltonian
  \be
 \Hp^\text{eff} (a,p_a;\bar{\phi},\bar{p}_\phi;\sigma)\equiv {\cal H}_G + \langle \Psi | \hat{{\cal H}}_{\phi} | \Psi \rangle. 
  \ee
 Semiclassical dynamics of the coupled matter gravity equations  arises from this effective  Hamiltonian by imposing the Poisson bracket 
$\{ \bar{\phi}, \bar{p}_\phi  \} =1$, which we henceforth write without the bars.  We define  $\hat{\cal H}_\phi$  in the polymer quantization described  below, and compute the expectation value.
 
\section{Polymer quantization of the scalar field}\label{sec:polymer}

Polymer quantization of the scalar field has been studied by several authors both at the formal level and applied to physical systems \cite{polyqm1,polyqft1,polyhyd,Kunstatter:2008qx}. There are two distinct versions of it depending on whether momenta or configuration variables are diagonal. Unlike in Schrodinger quantization there is no Fourier transform connecting these. The method we use  for polymer quantization on a curved background was introduced in \cite{polycosm1}.   

Let us consider the scalar field on the ADM background metric (\ref{admm}) and  the non-canonical phase space variables  \begin{equation}
\Phi_f \equiv \int d^3 x ~ \sqrt{q} f(x) \phi(x) , \quad U_{\lambda} \equiv \exp \left(\frac{i \lambda p_{\phi}}{\sqrt{q}} \right),
\end{equation}
where $f(x)$  is a smearing function. The parameter $\lambda$ is a spacetime constant with dimensions of $ (\text{mass})^{-2} $. These variables satisfy the Poisson algebra
\begin{equation}
\label{eq_pa}
\{ \Phi_f , U_{\lambda} \} = if\lambda U_{\lambda},
\end{equation}
Specializing  to  FRW spacetime with  line element
\begin{equation}\label{eq:FRW}
	ds^{2} = -N^{2}(t) \, dt^{2} + a^{2}(t) (dx^{2}+dy^{2}+dz^{2}), 
\end{equation}
we can set $f(x)=1$ because of homogeneity, so  these variables become 
\begin{equation}
\Phi = V_0 a^3 \phi , \,\,\,\, U_{\lambda} = \exp \left(\frac{i \lambda p_{\phi}}{a^3} \right),
\end{equation}
where $V_0$ is a fiducial volume. Their Poisson bracket is the same as in ($\ref{eq_pa}$). 

Quantization proceeds by realizing the Poisson algebra ($\ref{eq_pa}$) as a commutator algebra on the space of square integrable functions on the Bohr compactification of $\mathbb{R}$.  A basis is $ \{ \ket{\mu} ,\  \mu \in \mathds{R}  \} $ with the inner product
\begin{equation}
\braket{\mu^\prime}{\mu} = \delta_{\mu, \mu^{'}},
\end{equation}
where $\delta$ is the generalization of the Kronecker delta to the real numbers.  Basis wave functions  are $e^{ix\mu} = \langle x| \mu\rangle$ and the inner product in explicit form is 
\be
\langle \mu| \mu'\rangle \equiv \lim_{T\rightarrow \infty} \frac{1}{2T} \int_{-T}^T dx \ e^{-ix\mu} e^{ix \mu'} = \delta_{\mu,\mu'}.
\ee
  Action of the operators $ \hat{\Phi} $ and $ \hat{U}_{\lambda} $ are defined by 
\begin{equation}
\hat{\Phi}\ket{\mu} = \mu \ket{\mu}, \,\,\,\, \hat{U}_{\lambda} \ket{\mu} = \ket{\mu + \lambda}. \label{op-defs}
\end{equation}
A consequence of these operators  and the above inner product is that the momentum operator does not exist in this quantization because the translation operator is  not weakly continuous in $\lambda$. This may be seen by noting that the limit  
\be
\lim_{\gamma \rightarrow 0}  \frac{1}{\gamma} \langle \mu | U_\gamma - U_0    | \mu\rangle, 
\ee 
which could define the momentum operator using $U_\lambda$, does not exist. An alternative way to define the momentum operator  indirectly is  
\begin{equation}
\label{P_def}
p_{\phi}^{\lambda} \equiv \frac{a^3}{2i \lambda} ( U_{\lambda} - U_{\lambda}^{\dag} ), 
\end{equation}
a form  motivated by the classical expansion of  $U_\lambda$.  In particle systems  this modifies the kinetic energy operator. In LQG  it constitutes  the origin of ``holonomy corrections.'' The usual Schrodinger quantization results are obtained in suitably defined limits \cite{polyqmfredenhagen, Husain:2007bj, polyqmcorichi}.

For polymer corrected cosmological dynamics we wish to calculate the scalar field energy density  
\be 
\rho_{\text{eff}} = \frac{1}{a^3} \langle {\cal H}_\phi \rangle 
\ee
  for a  Gaussian coherent state peaked at the phase space values $(\phi , p_{\phi})$. Before doing so we fix the polymer energy scale by setting 
 $\lambda =\lambda_*\equiv M_\star^{-2}$. A suitable choice of state is 
 \begin{align} \nonumber
\ket{\psi} & = \frac{1}{\mathfrak{N}} \sum_{-\infty}^{\infty} c_{k} \ket{\mu_k}, \\ c_{k} & \equiv \exp \left[ - \frac{(\phi_{k} - \phi)^2}{2 \sigma^2} \right] \exp(- i p_{\phi} \phi_{k} V_{0}),
\end{align}
where $\phi_k \equiv \mu_k/V_0 a^3$ is an eigenvalue of the scalar field operator, rather than its ``integrated" version $\hat{\Phi}_k$ defined in (\ref{op-defs}) above.  

The effective density $\rho_{\text{eff}}$ was computed for the zero potential case in \cite{polycosm1}.  For completeness we summarize the main steps. The normalization constant is calculated by approximating the sum by an integral, 
\be
\mathfrak{N} = \sum |{c_{k}}|^2 \simeq V_{0} a^3 \sigma \sqrt{\pi},
\ee
and 
\be
\langle U_{\lambda*} \rangle =  e^{i\Theta} e^{-\Theta^2 /4 \Sigma^2},
\ee
where we have used the variables 
\be
\Theta \equiv \frac{p_\phi}{M^2_*a^3}, \ \ \ \Sigma= V_0\sigma p_\phi.
\ee
These are useful because they are invariant under the scale transformations  
\be
{\bf x}\rightarrow l {\bf x}, \ \  a\rightarrow l^{-1} a, \ \  V_0 \rightarrow l^3 V_0, \ \  p_\phi \rightarrow l^{-3} p_\phi.
\ee
Working with the pair $\Theta$ and $\Sigma$ ensures that there is no spurious dependence of results on $V_0$. Combining these equations leads to the effective energy density   
\be
\rho_{\text{eff}}(a,\phi,p_\phi;M_\star,\sigma) = \frac{M_\star^4}{4}\left(   1- e^{-\Theta^2/\Sigma^2} \cos 2\Theta   \right) + \frac{1}{2} m^{2} \left( \phi^{2} + \frac{\sigma^{2}}{2} \right).
\ee
This has an interesting limit: for $\Theta \gg 1$ and $\Sigma \approx 1$ the oscillatory term is damped out and the first term becomes an effective cosmological constant at the polymer scale.

\subsection{Polymer scalar equation of state}

 It is interesting to note the equation of state of the polymer quantized scalar field. Let us first define the dimensionless variables  
\bea
\vi &\equiv& \phi/\Mp, ~~ \pv \equiv \pp/\Mp^2, ~~ p_a \equiv P_a/\Mp^3 \nn \\
\gamma &\equiv& \Mp/\M, ~~ \delta \equiv m/\Mp, ~~ \lambda \equiv \Lambda/\Mp^2 \nn \\
\sigma &\rightarrow& \sigma/\Mp, ~~ V_0 \rightarrow V_0 \Mp^3, ~~ \H_p \rightarrow \H_p \Mp^4. \label{dvar}
\eea
 From (\ref{rhoeff}) the scalar field energy density (with zero cosmological constant)  becomes
 \be
  \rho_{\text{eff}} =   \frac{1}{4 \gamma^4}\left(   1- e^{-\Theta^2/\Sigma^2} \cos 2\Theta   \right)  
 + \frac{1}{2}   \delta^{2}\left(  \vi^{2} +     \frac{1}{2}  \sigma^{2}\right) \ , \label{rhoeff}
 \ee
 and the scalar field pressure   is 
 \be
 P =  \frac{1}{2\gamma^4} \exp{\left(-\frac{\Theta^2}{\Sigma^2}\right)} \Bigg[ \Theta \sin\left(2 \Theta \right) + \left( \frac{\Theta^2}{\Sigma^2} + \frac{1}{2} \right) \cos\left(2 \Theta \right) \Bigg] -\frac{1}{4\gamma^4} - \frac{1}{2}   \delta^{2}\left(  \vi^{2} +   \frac{1}{2}  \sigma^{2} \right),
 \ee 
 where $\Sigma$ is as defined above, and $\Theta = \pv \gamma^2/a^3$ in the dimensionless variables (\ref{dvar}) .
 
 In the limit of a small Universe, where $\Theta \rightarrow \infty$, these give
 \be
 \lim_{\Theta\rightarrow\infty}P =  - \lim_{\Theta\rightarrow\infty} \rho_{\text{eff}} =  -\frac{1}{4 \gamma^4}  - \frac{1}{2}   \delta^{2}\left(  \vi^{2} +     \frac{1}{2}  \sigma^{2}\right). \label{smalla}
 \ee
 On the other hand, in the limit of a large Universe where $\Theta \rightarrow 0$, 
 \bea
 P &=&   \frac{\Theta^2}{2\gamma^4} \left( 1+ \frac{1}{2\Sigma^2}   \right)  - \frac{1}{2}   \delta^{2}\left(  \vi^{2} +   \frac{1}{2}  \sigma^{2} \right) + \mathcal{O} (\Theta^4),  \label{Plate} \\
 \rho_{\text{eff}} &=& \frac{\Theta^2}{2\gamma^4} \left( 1+ \frac{1}{2\Sigma^2}   \right)  + \frac{1}{2}   \delta^{2}\left(  \vi^{2} +   \frac{1}{2}  \sigma^{2} \right) + \mathcal{O} (\Theta^4). \label{rholate}
 \eea 
  These are the classical results up to the polymer state width $\sigma$ corrections in the variable $\Sigma$, and in the potential. 
  The latter term is a cosmological constant-like  contribution to the energy density $\delta^2\sigma^2/4$, whereas the model we started with had zero cosmological constant. However this  does not  generate the vanishingly small  observed value without introducing correspondingly  small values of  $\delta$ and $\sigma$. Similarly, this term may be used to cancel a negative bare cosmological constant, but again this must be finely tuned.

\section{Semiclassical dynamics}\label{sec:EOMs}

Using (\ref{Hdust}) and (\ref{rhoeff}), the  effective Hamiltonian density is 
 \bea
\Hp^\text{eff} &=& \frac{p_a^2}{12 a}  - a^3 \frac{1}{4 \gamma^4}\left(   1- e^{-\Theta^2/\Sigma^2} \cos 2\Theta   \right) \nn \\
 &-& \frac{1}{2} a^3 \delta^{2} \vi^{2} - a^3 \left( \lambda + \frac{1}{4} \delta^2 \sigma^{2} \right),
\eea
and the corresponding equations of motion are   
\bea
\label{eq:EOM}
\dot{a} &=& \frac{1}{6} \frac{p_a}{a}\ ,  \\
\dot{p_a} &=& \frac{p_a^2}{12 a^2} +  3a^2\left(  \frac{1}{2}  \delta^2 \vi^2 +  \lambda + \frac{1}{4}  \delta^2 \sigma^2 + \frac{1}{4\gamma^4} \right)\ \nn   \\
&& - \frac{3}{2}\frac{a^2}{\gamma^4} \exp{\left(-\frac{\Theta^2}{\Sigma^2}\right)} \Bigg[ \Theta \sin\left(2 \Theta \right) + \left( \frac{\Theta^2}{\Sigma^2} + \frac{1}{2} \right) \cos\left(2 \Theta \right) \Bigg] \ , \label{padot}  \\
\dot{\vi} &=& -\frac{1}{2 \gamma^2} \exp{\left(-\frac{\Theta^2}{\Sigma^2}\right)} \sin\left(2 \Theta \right)\ ,   \\
\dot{\pv} &=& a^3 \delta^2 \vi\ .
\eea
As a standalone unconstrained Hamiltonian system, these equations are valid for all sets of initial data, including data for which $\Hp^\text{eff} =0$.  However, given their origin from dust time gauge, it is useful to understand the difference between the trajectories for zero and positive values of  $\Hp^\text{eff}$.  For this purpose only, there is no structural difference  in using the classical or semi-classical Hamiltonian, since these differ in only the  functional form of $\rho(p_\phi,a)$. Therefore let us use the classical  ${\cal H}_\phi$. 
 
 Writing the potential term as $V(\phi)$, the constant energy surface is then given by  
  \be
  \frac{e}{a^3} +V(\phi) = \left( \frac{p_a}{a^2} \right)^2 - \left(\frac{p_\phi}{a^3} \right)^2.
   \ee
A convenient set of variables to study this surface is  
$W^2\equiv 1/a^3$, $X^2\equiv V(\phi)$,  $Y \equiv p_a/a^2$,  $Z\equiv p_\phi/a^3$, so the energy surface is 
\be
eW^2 +X^2 = Y^2-Z^2.
\ee
Since $W,X,Y,Z$ are independent curvilinear coordinates on the phase space, it is clear that the $e=0$ energy surface has a Killing vector field $\p/\p W$. It is in this sense that the theory with dust ``reduces" to the  theory without dust in the full phase space. We will see in the numerical solutions below that the $e=0$ cases indeed reproduce the  results of the zero dust theory studied earlier in \cite{polycosm1,polycosm2}.

\begin{figure*}
\begin{center}
\includegraphics[width=5.5in,height=3.5in]{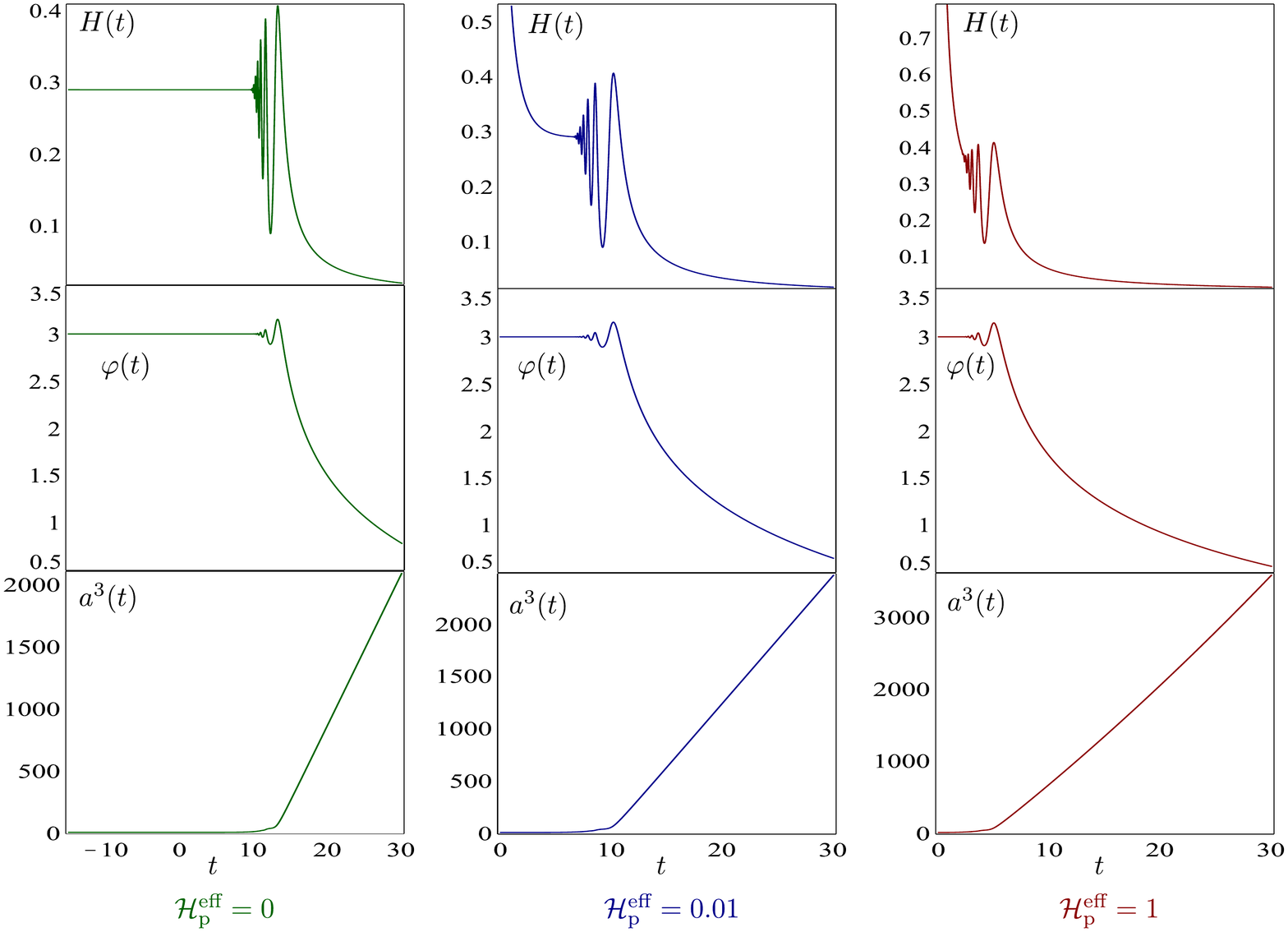}
\end{center}
\caption{Cosmological dynamics with massless scalar field for $\Hp^\text{eff} =0,0.01,1$. The graphs show the Hubble factor, scalar field and metric determinant as functions of dust time. There is a singularity at early time only for positive $\Hp^\text{eff}$. The reducing periods of polymer inflation as $\Hp^\text{eff}$ increases is evident.}\label{fig1}
\end{figure*}

\begin{figure*}
\begin{center}
\includegraphics[width=5.5in,height=3.5in]{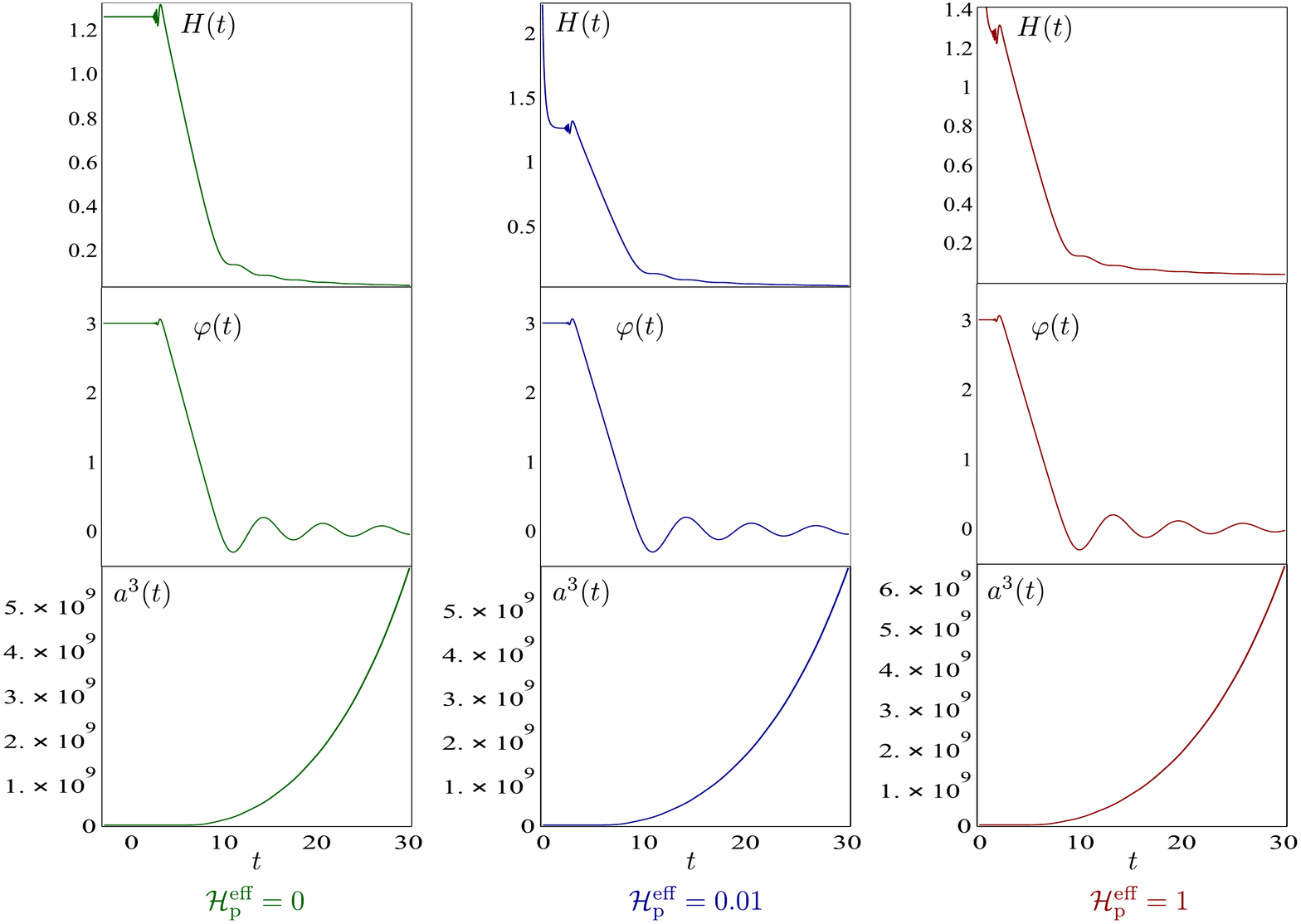}
\end{center}
\caption{Dynamics of a massive scalar field ($\delta=1$)  for $\Hp^\text{eff}=0,0.01,1$. The graphs show the Hubble factor, scalar field and metric determinant as functions of dust time. Polymer induced inflation is followed  by slow-roll and oscillatory decay, evident in the scalar field graphs.}\label{fig2}
\end{figure*}

\begin{figure*}
\begin{center}
\includegraphics[width=3.8in,height=2.9in]{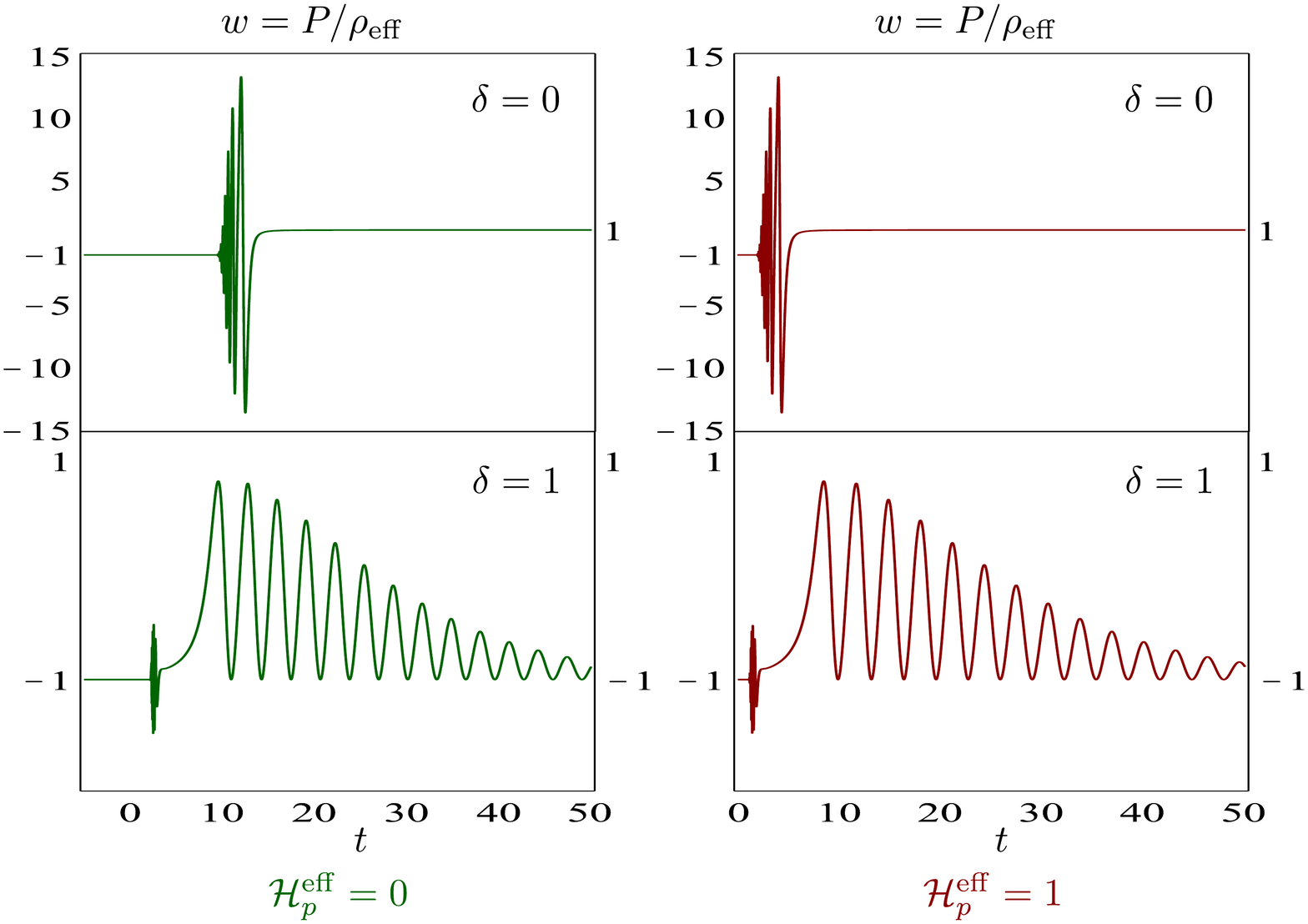}
\end{center}
\caption{The equation of state corresponding to two of the solutions in Figures \ref{fig1} and \ref{fig2}. The difference between the massive $\delta=1$ and massless $\delta=0$ cases is evident at late times.  }\label{fig3}
\end{figure*}

\subsection{Numerical results} 

We solved this system  of ordinary differential equations numerically using standard packages in MAPLE. For initial data such that  $\Hp^\text{eff}=0, 0.01, 1$ for  massless ($\delta=0$) and massive ($\delta=1$) scalars.  These cases illustrate the interesting change in dynamics caused by polymer quantization of matter. The results are in Figures 1 and 2. The other parameters  for  these illustrations are $\lambda=0, \gamma=1, V_0=1, \sigma=0.1$, and the initial data is $a(0)=0.1, \pv(0) = 100$,  $\vi(0)=3$. $p_a(0)$ is fixed by requiring the above values for $\Hp^\text{eff}$. 
 
  $ \Hp^\text{eff}=0.$
   This case appears in the left column in Figs 1 and 2. It reproduces  the result with no dust ($M=0$) studied in \cite{polycosm1,polycosm2}.    Although the above Hamiltonian equations of motion apply for all values of  $\Hp^\text{eff}$ including zero, it is useful to note that this case is degenerate in the sense that the constant energy surface has  a Killing vector field for $\Hp^\text{eff}=0$, as discussed above. 
    
  {$\Hp^\text{eff}>0.$}
    These cases have a curvature singularity. For zero scalar field mass, it is evident that after the singularity, there is a period of inflation that depends on the value of $\Hp^\text{eff}$. As in the $\Hp^\text{eff} =0$ case, this is  entirely a consequence of  polymer quantization of the scalar field and the unusual equation of state it produces. Larger values of $\Hp^\text{eff}$ give shorter durations of inflation. 

For non-zero scalar mass ($\delta=1$), shown in Fig. 2, there is a period of slow roll inflation following the polymer induced inflation. The oscillatory settling down to zero of the scalar field is also evident. As for the massless case, the period of polymer induced inflation decreases with increasing $\Hp^\text{eff}$. 

Another interesting feature in these plots  is that the scalar field does not diverge as $a^3(t)\rightarrow 0$, although there is a curvature singularity for $\Hp^\text{eff}>0$. This is clear from the scalar field equations, and has its origin in the semiclassical Friedman equation   
   \be
      3 \left( \frac{\dot{a}}{a} \right)^2 =   \frac{1}{\Mp^2} \left(\rho_\text{eff} + \frac{\Hp^\text{eff}}{a^3}\right) \ . \label{friedmann}
   \ee
From (\ref{rhoeff}), $\rho_\text{eff}$  is bounded above, so the singularity comes from positive values of  the constant $\Hp^\text{eff}$  on the right hand side. If $\Hp^\text{eff}=0$ the Hubble parameter too is bounded above,   and   there is eternal inflation and no singularity in the past.  
  
The numerical results presented are for a relatively small set of initial data parameters. But these are representative: all other values of parameters yield  results that are qualitatively similar.  The general feature is that larger values of $\Hp^\text{eff} =0$ give successively smaller periods of polymer quantization induced inflation. No significant effects occur  upon changes to the state width ($\sigma$) or mass $(\delta)$  parameters. 
 
Fig. 3 shows plots of the equations of state $w=P/\rho_{\text{eff}}$ computed for the same initial data as that used for Figs. 1 and 2. These demonstrate  another aspect of the polymer scaler field: the oscillations  between early and late times arise from the damped sinusoidal factors in the effective energy density and pressure. The massive case is qualitatively different due to the fact that there is mass and state width dependent effective cosmological constant $\Lambda\equiv \delta^2\sigma^2/4$ at late times, a feature that is clear from  eqns. (\ref{Plate}) and (\ref{rholate}).
 
   \section{Discussion}

We studied homogeneous and isotropic cosmology in a semiclassical  setting of polymer quantized matter and classical gravity. The approach is similar to the conventional  semiclassical Einstein equation, but it is carried out in the Hamiltonian  theory including back reaction.  We used the system of gravity coupled to scalar field and pressureless dust, and studied the dynamics using the physical Hamiltonian corresponding to dust time gauge. Our main result is that polymer quantization of the scalar field alone is responsible for a period of inflation after the initial singularity, a result we derived  using the  dust time gauge.   The number of e-folds of polymer quantization induced inflation is determined by the choice of initial data: smaller values of $\Hp^\text{eff}$ give more e-folds than larger values. 

It is useful to compare and contrast our results with those of Refs. \cite{polycosm1,polycosm2}, where  the semiclassical equations give an eternal period of inflation in the past. This is the case $\Hp^\text{eff}=0$, which we reproduced. The origin of the eternal inflation in the past  is  due to the fact that at early times the polymer Hamiltonian tends to a constant determined by the polymer scale, which acts like a cosmological constant.  In contrast, in the case $\Hp^{\text{eff}}=e>0$, there is an initial singularity followed by a period of inflation which depends on the value of $e$.  This is followed by a graceful exit into the classical regime. The reason there is a singularity for the $e>0$ case is of course that gravity is not quantized, and the Friedmann equation (\ref{friedmann})   contains the divergent term $e/a^3$. 

A possible method of ``resolving" this singularity while not fully  quantizing gravity sector is to employ a heuristic semi-classical  technique by replacing the $1/a^3$ factors in the scalar field energy density with a semiclassical expression  motivated by the Thiemann identity for defining the inverse triad operator \cite{Thiemann:1996aw}. This has been used in a number of works in cosmology \cite{Husain:2003ry,Singh:2005km} and in spherically symmetric gravitational collapse \cite{Husain:2008tc,Ziprick:2010vb,Husain:2008qx}. The net effect is that $1/a^3$ factors are replaced by a bounded function of $a$.  In the case of our physical Hamiltonian, these factors appear in the gravitational and scalar momentum parts of the Hamiltonian. This change would lead to  a significant modification of the dynamics in the region $a\rightarrow 0$. 

Although our approach is semiclassical in a specifically defined sense, and we use ADM variables, and an alternative polymer quantization not directly coming from LQG, we draw the lesson that  polymer quantization of matter in LQC models is likely to lead to interesting new dynamics. Our argument for this is that the main effect of  quantizing the gravity sector is to resolve the singularity -- this by itself should leave unaffected later time evolution, where the inflationary epoch is seen. But this temporal separation of  the singularity resolution time scale from the inflationary one might require distinct choices for the polymer scales for gravity  and matter, a question that requires further scrutiny in LQG. This may be studied in the LQC context by adding a scalar field to the gravity with dust model studied in \cite{dust-time-cosm}.

The main technical difference  in the LQC version of the model would be in polymer quantization of the matter sector, where the scalar field mass term  would be realized through an exponentiated operator $\displaystyle U_\phi\equiv e^{i\lambda \phi}$ \cite{polyqft1}. This of course has no effect in the theory without a potential. But with a mass term it would serve a role similar to that of the kinetic term in our version of quantization: the mass term would be realized as an operator made from the elementary operator $U_\phi$.

The role of dust as time may appear unappealing from a fundamental point of view, but it nevertheless provides a simple physical Hamiltonian which produces an inflationary phase in concert with polymer quantization of matter. Furthermore, the quantization method by itself has led to detailed predictions for density fluctuations \cite{Seahra:2012un}, which serves as a concrete potential test of the approach.

  \begin{acknowledgments}
  
We thank Sanjeev Seahra for discussions. This work was supported by the Natural Science and Engineering Research Council of Canada. S.M.H is supported by the Lewis Doctoral Fellowship.

\end{acknowledgments}

\bibliography{Polymer_dust_bib}

\end{document}